\documentclass[12pt]{article}
\usepackage{amsfonts}
\usepackage{a4wide}
\newcommand{\be}{\begin{equation}}
\newcommand{\ee}{\end{equation}}
\newcommand{\re}{{\rm e}}
\newcommand{\cZ}{{\cal Z}}

\newcommand{\fa}{\mathfrak{a}}
\newcommand{\fk}{\mathfrak{k}}

\newcommand{\Lie}{{\cal L}}
\newcommand{\vp}{\wedge}
\newcommand{\rD}{{\rm D}}
\newcommand{\CP}{\mathbb{C}\mathbb{P}}
\renewcommand{\P}{\mathbb{P}}
\newcommand{\C}{\mathbb{ C}}

\newcommand{\M}{\mathbb{ M}}
\newcommand{\tr}{{\rm tr}}
\newcommand{\ri}{{\rm i}}
\newcommand{\rd}{{\rm d}}
\newcommand{\p}{\partial}
\newcommand{\str}{\rule{0ex}{1ex}}
\newcommand{\half}{{\textstyle \frac{1}{2}}}
\newcommand{\cO}{{\cal O}}

\newtheorem{lemma}{Lemma}
\newtheorem{definition}{Definition}
\newtheorem{proposition}{Proposition}
\newenvironment{proof}{\medbreak \noindent {\bf Proof.} }{\medbreak}
\newcommand{\cA}{{\cal A}}
\newcommand{\cD}{{\cal D}}

\newcommand{\cH}{{\cal H}}

\newcommand{\cG}{{\cal G} }
\newcommand{\cM}{{\cal M} }

\newcommand{\fg}{\mathfrak {g} }
\newcommand{\fh}{\mathfrak {h} }

\newcommand{\gl}{{\rm gl}(N,\C)}
\newcommand{\lgl}{{\rm lgl}(N,\C)}
\newcommand{\GL}{{\rm GL}(N,\C)}
\newcommand{\LGL}{{\rm LGL}(N,\C)}
\newcommand{\LGLp}{{\rm LGL}_+(N,\C)}
\newcommand{\LGLm}{{\rm LGL}_-(N,\C)}
\newcommand{\lglm}{{\rm lgl}_-(N,\C)}
\newcommand{\LGLmone}{{\rm LGL}_{-1}(N,\C)}
\newcommand{\LGLmtwo}{{\rm LGL}_{-2}(N,\C)}

\newcommand{\sixth}{{\textstyle \frac{1}{6}}}

\begin{document}

\title{Duality for the general isomonodromy problem} \author{N.M.J
  Woodhouse\thanks{This work has been partially supported by the
    European Community (or European Union) through the FP6 Marie Curie
    RTN {\em ENIGMA} (Contract number MRTN-CT-2004-5652)}\\Mathematical
    Institute\\University of Oxford}

\maketitle
\begin{abstract}
By an extension of Harnad's and Dubrovin's `duality' constructions,
the general isomonodromy problem studied by Jimbo, Miwa, and Ueno is
equivalent to one in which the linear system of differential equations
has a regular singularity at the origin and an irregular singularity
at infinity (both resonant).  The paper looks at this dual formulation of
the problem from two points of view: the symplectic geometry of spaces
associated with the loop group of the general linear group, and a
generalization of the self-dual Yang-Mills equations.
\end{abstract}

\subsection*{Introduction}

The isomonodromy problem studied by Jimbo {\em et al} (1981a,b,c) concerns the
deformations of a linear systems of ordinary differential equations in
the complex plane of the form
\be
\frac{\rd y}{\rd x}= Ay\, .\label{sys}
\ee
Here $x\in\C$, 
$y$ is column vector of length $\ell$, and $$
A(x)= \frac{Q(x)}{q(x)}
$$
is an $\ell \times \ell $ matrix-valued rational function of $x$,
with $q$ is complex polynomial, $Q$ is matrix-valued polynomial,
and $\deg (Q)=\deg( q)-2$.  The poles of $A$ are the roots $\lambda_1,
\ldots , \lambda_p$ of $q(x)$, with the order of the $i$th pole equal
to the multiplicity $k_i$ of $\lambda_i$.  
With Jimbo {\em et al}, we assume that $Q(\lambda_i)$ has
distinct nonzero eigenvalues for each $i$.

By imposing the condition on the degree of $Q$, we diverge from Jimbo
{\em et al} in excluding singularities in the system at $x=\infty$;
but as the theory is invariant under conformal transformations of $x$,
this restriction is not a real one. We can always move a singularity
at infinity to a finite value of $x$ by making a M\"obius
transformation.

The {\em isomonodromic deformation problem} is to determine variations
of $A$ that preserve the {\em monodromy data} of the system---that is,
the monodromy representation together with the Stokes' matrices at the
irregular singularities and the connection matrices between the
solutions with particular asymptotic behaviour at the poles.  A
solution is a family of ODEs of the form (\ref{sys}) with the same
data, parametrized by local coordinates on a {\em deformation
  manifold} $\cD$.  Jimbo {\em et al} characterize the isomonodromic
dependence of $A$ on the coordinates in terms of the existence of a
matrix-valued 1-form $\Omega$ with rational functions of $x$ as
entries.  The properties of $\Omega$ are determined by the
singularities in $A$; in particular, $\Omega$ has poles in $x$ only at
the singularities of $A$. The central condition is that the curvature
of the meromorphic connection
$$
\nabla=\rd- A\, \rd x +\Omega
$$
should vanish.

In this paper, I shall explore some aspects of another formulation of
the problem.  The starting point is that deformation conditions
can be be rewritten in the form
\be
\rd b +[\beta, b]=0, \qquad [\beta,a]=[b,\rd a].
\label{secform}\ee
Here $a,b$ are $N\times N$ matrices, where $N=\ell\deg{q}$.  We think
of $a$ as the independent variable and $b$ as the dependent variable.
The independent variable $a$ takes values in an abelian subalgebra
$\fa \subset \gl$ determined by the original problem, and $\alpha$ and
$\beta$ are matrix-valued 1-forms on $\fa$, regarded as a complex
manifold. The generic matrices in $\fa$ all have the same Jordan
canonical form, with eigenvalues determined by the positions of the
poles of $A$ and with the sizes of their Jordan blocks determined by
their orders.  The second equation determines $\beta$, up to a
residual gauge freedom.  The first then determines the dependence of
$b$ on $a$.  The matrix $b$ must satisfy three constraints:
\begin{itemize}
\item it has rank $\ell$ and $b^2=0$; 

\item it satisfies an algebraic
constraint that ensures that the second equation can be solved for
$\beta$; 
\item  with $E$ denoting the {\em Euler vector field} that generates
the flow $a\mapsto \re^{\tau}a$, $\tau\in\C$, 
we have  $b=i_E\beta$ and $\Lie_E\beta=0$.
\end{itemize}
The constraints are compatible with (\ref{secform}).

The relationship between the two forms of the problem is an extension
of Harnad's duality (1994).  Indeed (\ref{secform}) implies that
the deformations of the system
$$
z\frac{\rd v}{\rd z}+(a z + b)v=0\,,
$$
are isomonodromic as $a$
varies. However this is a nonstandard isomonodromy
problem because $a$ is not diagonalizable: not only does it have a
nontrivial Jordan canonical from, but it also has $\ell$ Jordan blocks
corresponding to each eigenvalue. Both these features lead to
complications that are not immediately apparent in the simple form of
the equations (\ref{secform}). The general theory of deformations of
such `resonant' singularities has been explored recently by Bertola
and Mo (2005).

Given a solution of (\ref{secform}), one can recover a solution of the
original problem by noting that (\ref{secform}) implies that the
connection form
$$
\rd -(a-x)^{-1} \,(\rd a-\rd x) \,b +\beta
$$
has zero curvature.  It induces a flat meromorphic connection on
the bundle $\C^N/\ker b$ over the deformation manifold, 
which coincides up to gauge with $\nabla$.

The second form of the deformation equations leads to two other ways
of looking at the problem.  The first is in terms of the symplectic
geometry of a manifold constructed from the loop group $\LGL$ by taking
some ideas from
Pressley and Segal (1986), pp
49--50. We let $\Gamma\subset\C$ denote 
unit circle in the complex plane and consider the set of smooth
1-forms on $\Gamma$ with values in $\gl$.
Each 1-form $\mu=B\,\rd 
z$
determines a monodromy matrix $m$, up to conjugacy: it is defined by picking a
fundamental solution $y:\Gamma\to\GL$ of the differential equation
\be
\rd y+\mu y=0
\label{diffeqn}\ee
and by taking $m$ to be the constant matrix $y^{-1}\tilde y$,
where $\tilde y$ is the continuation of $y$ once around $\Gamma$ in
the positive sense.
Since $y$ is
unique up to multiplication on the right by a constant matrix, the
monodromy is determined by $\mu$ up to conjugacy.

We denote by $\cM$ the set of 1-forms whose monodromy is conjugate to
a fixed matrix $m$.  Any two elements $\mu, \hat \mu$
of $\cM$ are related by a gauge transformation
$$
\hat \mu= g^{-1}\mu g+g^{-1}\rd g \,.
$$
A tangent to $\cM$ is determined by a map $h:\Gamma \to\gl$ by
$$
\delta\mu=\rD h= \rd h +[\mu,h]\,.
$$
So we can define a
symplectic form on $\cM$ by
$$
\omega(h,h')=\frac{1}{2\pi\ri}\oint\tr (h\rD h')\, .
$$
It is skew symmetric, by integration by parts, and non-degenerate
since $\omega (h, \,.\,)=0$ only if $\rd h -\mu h= 0$.  It is also
closed.

By using Birkhoff's factorization theorem, almost every $\mu\in \cM$
can be reduced by a gauge transformation by $f_-\in\LGLm$ to
the form
\be
\mu_+ +z^{-1} b \,\rd z
\label{redform}\ee
where $b$ is constant, 
with $\exp(2\pi\ri b)$ conjugate to $m$, and where $\mu_+$ extends holomorphically to the
interior of $\Gamma$ in the complex plane.  The
gauge transformation 
is uniquely determined by $\mu$ if we impose the condition
$f_-(\infty)=1$.

Within this framework, we obtain a simple symplectic interpretation of
the deformation equations.  We consider the group ${\rm
  LGL}_{-2}(N,\C)$ of loops that extend holomorphically to the outside
of $\Gamma$ and are of the form $1+O(z^{-2})$ as $z\to \infty$.  This
group has a Hamiltonian action on $\cM$ by gauge transformations and
the resulting Marsden-Weinstein reduction is a finite-dimensional
complex symplectic manifold.  For points of this manifold, the gauge
transformation to (\ref{redform}) results in a 1-form
$$
(a+z^{-1}b)\, \rd z\,.
$$
Thus we have a projection from the reduced symplectic manifold  
onto systems of the form
$$
\frac{\rd v}{\rd z}+(a  + z^{-1}b)v=0\,.
$$
The Hamiltonians are defined from spectral invariants of $B$ in a
neighbourhood of $z=\infty$. They of two types: Hamiltonians in
involution that move the poles of $A$, but leave the singularity
behaviour at the poles unchanged but for a coordinate transformation;
and others that fix the poles, but change the remaining deformation
parameters. The flows commute with the projection, and generate
isomonodromic deformations.

The final results concern the interpretation of (\ref{secform}) as a
symmetry reduction of a generalized form of the self-dual Yang Mills
equations on a `space-time' which is the product $\M=\fa\times\fa$ of
two copies of $\fa$.  The space-time variables are matrices $s,t\in
\fa$.  Given a solution to the deformation problem, we put \be
\rD=\rd+\Phi=\rd+\rd s+ \beta (t)\,\qquad (s,t)\in \M\,,
\label{sdconn}\ee
and regard $\rD$ as a connection on the trivial $\C^N$ bundle
over $\M$.  The curvature of $\Phi$ is
$$
F=\rd \beta + \beta\vp\beta + \rd s \vp \beta + \beta\vp \rd s.
$$
If we restrict the connection to an $N$-plane in $\M$ of the form
\be
\omega= \pi_1s+\pi_0t
\label{nullplane}\ee
where $\omega\in \fa$ and  $(\pi_0,\pi_1)\in \C^2$ are constant,
then the curvature of the restricted connection is
$$
\rd \beta +\beta\vp\beta 
-z(\rd t \vp \beta + \beta\vp \rd t)\,,
$$
where $z=\pi_0/\pi_1$. This vanishes as a consequence of the deformation
equations (\ref{secform}).  Conversely, if 
$F$ vanishes on every such $N$-plane and $\Lie_E\beta=0$, then
(\ref{secform}) holds with $b=i_E\beta$.

There is a close analogy with the condition that a connection on
complex Minkowski space should be a solution to the self-dual Yang-Mills
equations, namely that the restriction of the connection to null
$\alpha$-planes should be flat (Ward and Wells 1990, p.\ 373).  Indeed
when $\fa=\C^2$ as a vector space, it is precisely that.  So we call
the $N$-planes defined by (\ref{nullplane}) for constant $\omega$ and
$\pi_A$ {\em null $N$-planes} and we say that a connection
$\rD$ that has vanishing curvature on every null $N$-plane is {\em
  self-dual}. The terminology is derived purely from the analogy: there is no
metric on the `space-time' with respect to which the `null $N$-planes'
are null, and no duality operator on 2-forms.

The `self-duality' condition involves only the structure of $\fa$ as a
vector space. It has been studied as a generalization of the self-dual
Yang-Mills equation (see Mason and Woodhouse 1996, \S8.6).  In itself,
it does not involve the algebraic structure of $\fa$. This comes into
play when we look at the extra invariance conditions on $\rD$ that
follow from the special form of (\ref{sdconn}).  The transformations
which are of interest here are those that map null $N$-planes to null
$N$-planes.  These include the three flows
$$
(s,t)\mapsto (s,\re^{\tau}t), \qquad
(s,t)\mapsto (s+\tau c, t), \qquad
(s,t)\mapsto (s, t+\tau c)\,,
$$
where $\tau\in\C$ is the parameter along the flows and $c\in\fa$ is constant.
We denote the respective generating vector fields on $\M$ by $E$,
$S_c$, and $T_c$.   The connection form $\Phi$ in (\ref{sdconn})
has vanishing Lie derivative along the `Euler
vector field' and along the vector fields $S_c$, $c\in\fa$.

In a general setting, a connection form $\Phi$ 
can undergo a {\em gauge transformation}
$$\Phi\mapsto g^{-1}\Phi g+g^{-1}\rd g\,,$$ where $g$ takes values in the
corresponding gauge group.  It is {\em equivariant} along a vector field $X$
if $g$ can be chosen so that $\Phi$ is invariant in the sense that its
Lie derivative along $X$ vanishes.  In this case the
corresponding {\em Higgs field} $\phi_X=i_X\Phi$ transforms by
conjugation under gauge transformations that preserve the
invariance. 

With this terminology,  solving  the deformation problem is
equivalent to finding a `self-dual' $\GL$ connection on $\M$
satisfying the invariance conditions:
\begin{itemize}
\item  It is equivariant along $E$ with Higgs field conjugate to the
  monodromy generator.

\item It is equivariant under the translations $S_c$, and the image of
  $\fa$ under $c\mapsto \phi_{S_c}$ is everywhere conjugate to $\fa$
  in $\gl$.
\end{itemize}
Note that if $\Phi$ is equivariant under the Lie algebra of
translations in $s$, then the image is in any case necessarily
abelian.  The structure of $\M$ allows to us define a linear operator
on 1-forms on $\M$ by $\alpha\mapsto \alpha^{*}$, 
$$
i^{\str}_{T_c}\alpha^{*}=i^{\str}_{S_c}\alpha, \qquad
i^{\str}_{T_c}\alpha=i^{\str}_{S_c}\alpha^{*}\,.
$$
In this notation, 
a general gauge potential can be written $\Phi=\alpha^{*}+\beta$, where
$\alpha$ and $\beta$ contain only $\rd t$ terms.
Under the invariance conditions, we can choose the gauge so that the
components of $\alpha$ and $\beta$ are functions of $t$ alone,
$\alpha$ takes values in $\fa$, and 
\be
\Lie_X\alpha =\alpha, \qquad \Lie_E\beta=0\,.
\label{LieE}\ee
The self-duality condition is
$$
\rd\alpha +\alpha
\vp\beta +\beta\vp\alpha=0, \qquad \rd \beta +\beta\vp \beta=0\,.
$$
Proposition \ref{appprop} in Appendix B then implies that the gauge
can be chosen so that $\rd\alpha=0$ and so that (\ref{LieE})
also holds in the new gauge. If we put $a=i_E\alpha$, we then  have
$$
\alpha=\rd a, \qquad \rd a\vp\beta +\beta\vp\rd a=0, \qquad
  \rd \beta +\beta\vp \beta=0\,.
$$
By changing the $t$-variable to $a$, we then have a connection in the
form
(\ref{sdconn}). 
With $b=i_E\beta$, we also have (\ref{secform}). In fact, under the
invariance conditions, the self-duality condition is equivalent to
(\ref{secform}), with $\Lie_E\beta=0$, by this construction.

An application of standard twistor methods gives the following
result, which can be understood as an explanation, in part, of the
Painlev\'e property of the deformation equations (see also Beals and
Sattinger 1993).  Suppose that $b(a)$ is a solution to (\ref{secform})
on a simply-connected open set $W\subset \fa$, with $b=i_E\beta$ and
$\Lie_E\beta=0$.  Then the linear system
$$
\rd f +\beta(a)f + z \rd a f=0
$$
has a solution $f(a,z)$ on $W\times \C$ that depends
holomorphically on $z$.  Suppose that $s,t\in\fa$ are such that 
$$ 
t+ \re^{\ri\theta} s\in W
$$
for all real $\theta$.  Then for generic $s$ and $t$ satisfying this condition , we have a unique 
Birkhoff factorization
$$
f(z^{-1} s +t,z)= f_-^{-1}(s,t,z)f_+(s,t,z)
$$
where $f_+$ and $f_-$ holomorphic with respect to $z$, respectively
inside and outside the unit circle in the $z$-plane, and $f_-=\exp(s)$
at $z=\infty$. We prove the following.
\begin{proposition}
The value of $f_+$ at $z=0$ is independent of $s$, and
$f_+(s,t,0)=f(t,0)$ for $t\in W$.
\end{proposition}
Since $f(a,0)$ determines $\beta$ and hence $b=i_E\beta$, the 
effect is to reduce the problem of propagating $b$ out of $W$ to the
solution of a Riemann-Hilbert problem. Singularities arise at the
points where the factorization fails, but they are poles.

\subsection*{Isomonodromic deformations}

Suppose that $\lambda_i$ is a root of $q$ with multiplicity $k_i$, 
and put $x_i=x-\lambda_i$.  By diagonalizing $Q$ in a
neighbourhood of $\lambda_i$, we can write
\be
A= g_i(x_i)\frac{\rd}{\rd x_i}
\Bigl(x_i^{-k_i+1}t_{i}(x_i)+m_{i}\log x_i\Bigr)g_i(x_i)^{-1} +O(1)
\label{aform}\ee
as $x_i \to 0$, where $m_i$ is a constant diagonal matrix and
$t_i(x_i)$ is a diagonal matrix with entries polynomial of degree
$k_i-2$ in $x_i$. At the regular singularities ($k_i=1$), we put
$t_{i}=0$.  The diagonal entries in $m_i$ are called an {\em exponent of formal
  monodromy}. 

The matrices that diagonalize  $A$ near the poles, and hence determine its
singular parts, need only be known up order
$x_i^{k_i}$.  They coincide up to this order with the matrices that
are found in the analysis in Jimbo {\em et al} as formal power series
in $x_i$ by looking for formal series solutions to the linear system
in a neighbourhood of each pole.  

The exponents of formal monodromy are unchanged in the
deformations, which are parametrized by the positions of the poles and
by the coefficients of the polynomials $t_{i}$ at the irregular
singularities.  The parameters are local coordinates on a complex {\em
  deformation manifold} $\cD$ of dimension $\sum_i^p(\ell (k_i-1)+1)$.

The infinitesimal isomonodromic deformations of $A$ are determined by 
the 1-form
\be
\Omega
=\sum_i\bigl(A' \rd \lambda_i -  
x_i^{-k_i+1}g_i\rd t_i g_i^{-1}\bigr)_{i,-}\,\,,
\label{Omform}\ee
where $\rd$ is the holomorphic exterior derivative on $\cD$, with $x_i$
held constant, the prime denotes differentiation with respect to
$x_i$, and the subscript $i,-$ denotes the negative terms in the Laurent
expansion in $x_i$. 
The components of $\Omega$ are matrices with
rational functions of $x$ as entries. 
Jimbo {\em et al} 
show that solving  isomonodromic deformation problem is equivalent to satisfying the
condition that $$\nabla=\rd+ \Omega-A\,\rd x $$ should be a flat meromorphic
connection.  That is, \be\rd A= [A,\Omega]-\frac{\rd\Omega}{\rd x}, \qquad 
\rd \Omega+\Omega\vp\Omega=0\,.
\label{defeqn}\ee

The flatness condition is preserved by transformations
$$
A\mapsto g^{-1}Ag, \qquad \Omega \mapsto g^{-1}\Omega g +g^{-1}\, \rd g \,,
$$
where $\rd$ is the exterior derivative on $\cD$ and $g:\cD\to {\rm
  GL}(\ell,\C)$; that is, by gauge transformations independent of $x$.
Thus we should think of the connection as being defined on the
pull-back by the projection $\pi\::\:\cD\times \CP_1\to \cD$ of a
holomorphic vector bundle $E\to \cD$. 
With $\Omega$ is defined by
(\ref{Omform}), we have that
$\Omega_{\infty}=\Omega\vert_{x=\infty}=0$. 
In a general gauge,
however, $\Omega_{\infty}\neq 0$.  

We denote by $\nabla_{\infty}=\rd
+\Omega_{\infty}$ the flat connection on $E$ given by restricting
$\nabla$ to $z=\infty$.
More generally, $\rd +\Omega_{\infty}$ can be any flat connection on $E$.
\begin{definition} By a {\em meromorphic $q$-connection} $\nabla$ 
on $\pi^*E$ is 
meant a connection represented locally by a $1$-form 
$$
\Omega- q^{-1}(x)Q(x)\,\rd x
$$
where $\Omega$ 
has poles in $x$ coinciding in location and multiplicity with the
roots of $q$ and at each root $\lambda_i$,
$ q(\lambda_i)\Omega(\lambda_i)= Q(\lambda_i)\,\rd \lambda_i$.  If in
addition $\nabla$ is flat, then it is said to be a {\em solution of
  the deformation problem}.
\end{definition}

Eqn (\ref{defeqn}) determines the isomonodromic deformations of $A$:
given $A$ at one point of $\cD$, we can compute $\Omega$, and so
determine the deformed linear system at a nearby point.  A
straightforward, but intricate, calculation turns this into a system
of nonlinear equations for the coefficients of $Q$ as functions on
$\cD$.  Our alternative form of the deformation equations uses a
different coordinate system on $\cD$, in which the entries in $a$
replace the parameters $\lambda_i$ and $t_i$.  The explicit
relationship is given in Appendix A.

\subsection*{Structure of the deformation manifold}

The tangent space to the deformation manifold at each point has the
structure of a finite-dimensional complex commutative algebra  with
identity, with multiplication law defined as follows.  For a
tangent $X$ to $\cD$, put $\Omega_X=i_X\Omega$.  If $X,Y$ are tangent at
a point, then we define $XY$ by the condition that
$$
\Omega_{XY} -\Omega_XA^{-1}\Omega_Y
$$
should be holomorphic at the poles. The
identity $I$ is the deformation determined by $\Omega_I=A$
(translation of the poles by the same constant,
leaving the other parameters fixed).
 
To put this another way, let
$\fh_i$ denote 
the quotient of the algebra holomorphic maps from a
neighbourhood of the pole $\lambda_i$ into $M_{\ell}(\C)$ 
(the $\ell\times \ell$
complex matrices) by the ideal generated by $q(x) 1_{\ell}$. 
Let
$$
\fh= \bigoplus_i\fh_i
$$
Then the first $k_i$ terms in the Taylor expansion of
$A^{-1}\Omega$ at each $\lambda_i$ determine a 1-form $\alpha_i$ on
$\cD$ with values in $\fh_i$. The sum $\alpha=\sum_i\alpha_i$ is a
1-form with values in $\fh$ and the algebraic structure in the
tangent space is induced by matrix multiplication of the $\alpha_i$s.
The fact that the image of the tangent space under $\alpha$ is a
commutative subalgebra of $\fh$ follows from the form of $A$ and
$\Omega$ (\ref{Omform}). The algebraic structure
can be expressed entirely in terms of the coordinates on $\cD$,
independently of the particular solution of the deformation
problem.  We shall see below that the deformation equations take a
particularly simple form in coordinates that are well adapted to the
algebraic structure.

The tangent space at a point of $\cD$ 
decomposes as a vector space into a sum
of the semi-simple subalgebra  spanned
by the deformations for which
$$
i_X\alpha_i=X(\lambda_i)1_{\ell}
$$
and the subalgebra made up of the nilpotent elements, which are
characterized by the vanishing of $i_X(A^{-1}\Omega)$ at the poles.
The former deformations simply move the poles without changing the
singular parts of $A\,\rd x$, beyond a coordinate transformation. The
latter fix the poles, but can change the other deformation parameters.

The original system of ordinary differential equations is determined
by the 1-form $A\, \rd x$ on the Riemann sphere.  There is a special
class of deformations which are given by the action of ${\rm SL}(2,\C)$
on $\CP_1$. They are given by
$$
\Omega_X= c(x)A\,,
$$
where $c$ is a complex quadratic in $x$.  Two important special cases are
$c=1$, where the corresponding vector field $I$ on $\cD$ translates all
the poles, leaving $A$ otherwise unchanged; and the {\em Euler vector
field} $E$, given by $c=x$, which rescales the coordinate $x$.  The
vector field $I$ is the identity in the tangent space at each point.
That is, $IX=X$ for any tangent $X$ to $\cD$.

\subsection*{The transformed problem}

The general isomonodromy problem can be transformed into a simpler one
in which the linear system of ODEs is 
\be
z\frac{\rd v}{\rd z}+(a z + b)v=0\,,
\label{translinsys}\ee
where $v$ takes values in a vector space $V$, and $a$ and $b$ are
elements of the Lie algebra of the endomorphism group of $V$. This
system has a regular singularity at $z=0$ and an irregular one at
infinity.  The corresponding isomonodromy problem is familiar in a
number of contexts in the case that $a$ is diagonalizable---the
non-resonant case.  For example, in the theory of Frobenius manifolds
(Dubrovin 1996, Hitchin 1997).  But in the problem that we consider, $a$ has
nontrivial Jordan canonical form, with the locations of the poles in
the original problem as eigenvalues, and block structure determined by
their ranks. There is the further complication that there are $\ell$
Jordan blocks corresponding to each eigenvalue.

The vector space $V$ in the transformed problem is defined to be the
direct sum $$V=V_1\oplus \cdots\oplus V_p\,,$$where $V_i$ is the space of
complex polynomials in the variable $x_i$ of degree $k_i-1$, with
values in $\C^{\ell}$.  Thus $v\in V$ is an $p$-tuple $(v_1, \ldots ,
v_p)$, where $v_i$ is a polynomial in $x_i$, taking values in
$\C^{\ell}$.  The dimension of the whole space is $N=\ell\sum k_i$.

Given the $\lambda_i$s, we can put $x_i=z-\lambda_i$, and so represent
the elements of $V$ in terms of equivalence classes holomorphic maps
$u:U\to \C^{\ell}$, where $U$ is the union of disjoint open
neighbourhoods $U_i$ of the points $\lambda_i$ on the Riemann sphere.
Two maps $u,u'$ are equivalent whenever their difference has a zero of
order $k_i$ at $\lambda_i$ for each $i$. We go back and forth between
the two representations by putting $x_i=x-\lambda_i$ and by defining
$v_i$ to be the first $k_i$ terms in the expansion of $u$ in powers of
$x_i$ at $x=\lambda_i$.

We denote the endomorphism group of $V$ by $\cG$ and its Lie algebra
by $\fg$.  A linear transformation $g\in\cG$ is a block matrix \be
g=\pmatrix{g_{11}& g_{12} & \ldots &g_{1p}\cr g_{21} &g_{22}&
  \ldots&g_{2p}\cr \vdots&&&\vdots\cr g_{p1} &g_{p2}& \ldots&g_{pp}\cr}\,,
\label{gblock}\ee
where $g_{ij}:V_j\to V_i$. 
A key subgroup $\cH\subset \cG$ is the group of
transformations $u\mapsto hu$, where $h:U\to {\rm
GL}(\ell,\C)$ is holomorphic; two such maps give the same element
of $\cG$ whenever their difference is a holomorphic multiple of $q$.  
In this subgroup, the the off-diagonal blocks (\ref{gblock}) are
zero.  
If we construct a basis for $V$ by using the coefficients of the
$v_i$s as linear coordinates, ordered appropriately,
then the diagonal blocks  are themselves
block matrices of the form
\be \pmatrix{ h_0 & h_{1} &h_{2} 
  &\ldots& h_{k_i-2}
& h_{k_i-1}\cr 0& h_0 & h_{1} & \ldots &
h_{k_i-3}&  h_{k_i-2}\cr \cr &&\ddots &\ddots &&\cr 
\cr
0&0&0& \ldots&h_0 &h_{1}\cr
0&0&0& \ldots &0&h_0\cr}\, ,
\label{alphblock}\ee
where the $h_j$s are $\ell\times\ell$ matrices---the coefficients in
the expansion of $h(z)$ around the corresponding pole. We shall go
back and forth between two interpretations of an element $h$ of $\cH$,
or of an element of the Lie algebra $\fh$ of $\cH$: as a linear map
$V\to V$ and as a holomorphic map $U\to {\rm GL}(\ell,\C)$.

Within $\cH$, there is the abelian subgroup $\cA$, in which $h$ is
diagonal and $h(\lambda_i)$ is a scalar multiple of the identity for
each $i$. Here the $h_i$s are diagonal and the $h_0$s are multiples of
the identity. The Lie algebra $\fa$ of $\cA$ will be central in what
follows. It is the model for the algebraic structure on the tangent
space at each point of the deformation manifold.

In passing from the original isomonodromy problem to the new one, we
define $a$ and $b$ by
\be
au=xu, \qquad bu=\sum \frac{1}{2\pi\ri}\oint \nabla u
\label{abdefn}\ee
where the integrals are around contours surrounding the poles and the
sum is over the poles.  Thus $bu$ is a constant map $U\to \C^{\ell}$.
In the  basis above, $a$ is in Jordan canonical form. If $h\in\fg$
and$[h,a]=0$, then $h\in \fh$.

\begin{lemma}
Suppose that $h\in \fh$ and $b(a-w)^{-1}hv=0$
  identically in $w\not\in U$ for every constant $v\in V$.
 Then $h=0$.
\label{vanishing}\end{lemma}
\begin{proof}  
The representation of $h$ as a matrix-valued function of
$x\in U$ is unique up to the addition of a
holomorphic multiple of $q(x)$.  In particular we can define $h(x)$ by
$$
h(x)v=q(x)\sum\frac{1}{2\pi\ri}\oint \frac{(hv)(w)\,\rd w}{(w-x)q(w)}\,
$$
for constant $v$.
We then have that, for any constant $v$, 
$$
b(a-x)^{-1}hv=\sum\frac{1}{2\pi\ri}\oint \frac{Q(w)h(w)v\,\rd
  w}{(w-x)q(w)}
=A(x)h(x)v\,.
$$
So if the left-hand side vanishes identically in $x$, then $h(x)=0$.
\end{proof}

\subsection*{Deformations of the transformed system}

We want to understand the  isomonodromic deformations of
(\ref{translinsys}) when $b\in\fg$ and 
$$
a\,:\, u\mapsto xu\,.
$$
We shall show that they equivalent to the deformations of the original
problem: the relationship is a variant of Harnad's (1994) duality, and
of the Dubrovin's (1996) transformation between two forms of the
equations for a Frobenius manifold. The complication is that $a$ is
not diagonalizable, except in the case that the original system has
only regular singularities, so that we cannot apply directly the
standard theory.

In the basis defined above, $a$ is in Jordan canonical form. Its
off-diagonal block are zero, and the diagonal blocks are
$$
\pmatrix{ \lambda_i & 1_{\ell} &0 & 0 &\ldots &
  0\cr
0&  \lambda_i & 1_{\ell} &0 & \ldots &
  0\cr
&& \ddots &\ddots &&&\cr
0&0&0&0& \ldots & \lambda_i\cr           }\,,
$$
We shall be interested only in deformations that leave the
positions of the poles at $0$ and $\infty$ fixed. It is not immediately apparent what is
meant by `isomonodromic deformation' in this setting.  In passing from
the non-resonant case, we take the first of eqns (\ref{defeqn}) to
characterize `isomonodromy'. Since the system has a regular
singularity at the origin and a pole of order two at infinity,
 we look for deformations of the form
$$
\delta(a+z^{-1}b) = \frac{\rd\omega}{\rd z} +[a+z^{-1} b, \omega]
$$
where $\omega=z a'+ b'$, with $a',b'\in\fg$. The condition
that the deformed linear system should be of the same form as the original 
implies
\be
[a,a']=0, \qquad [a,b']+[b,a']=\delta a-a'\, ,
\label{alphbet}\ee
where $\delta a$ is the variation of the matrix of $a$: it is diagonal
with diagonal entries $\delta \lambda_i$.

When $a$ is diagonal with distinct eigenvalues, the first equation
is satisfied whenever $a'$ is also diagonal; the second can then
be solved for $b'$, up to the addition of a diagonal matrix.  In
our case, however, the first equation implies that $a'$ lies in $\fh$,
but the second imposes algebraic further constraints on $a'$.

To understand how such deformations arise from a solution to the
original problem, we need some notation.  The elements of the Lie
algebra of $\cH\subset\cG$ are holomorphic maps $h:U\to {\rm
  gl}(\ell,\C)$, modulo the addition of maps that vanish to order
$k_i$ at each $\lambda_i$.  We define a nondegenerate bilinear form on
$\fh$ by
$$
\langle h,\hat h\rangle = \frac{1}{2\pi\ri}\sum_i \oint \tr \bigl(x^{-k_i}
h(x)\hat h(x)\bigr)\rd x\,,
$$
where the integrals are around the poles $\lambda_i$.  We then have a
linear projection
$\fg \to\fh:b\to h_b$ defined by
$$
\langle h_b,h\rangle = \tr(bh)\qquad\mbox{for $h\in \fh$}
$$
where on the right the trace is in $\fg$.
Suppose  we write $b$ in the block form (\ref{gblock}). For each
$i$, the $i$th
diagonal block can be written in the basis above as
\be
\pmatrix{b^{i}_{11}&b^{i}_{12} & \ldots &b^{i}_{1k_i}\cr
b^{i}_{21}&b^{i}_{22} & \ldots &b^{i}_{2k_i}\cr
\vdots&&&\cr
b^{i}_{k_i1}&b^{i}_{k_i2} & \ldots &b^{i}_{k_ik_i}\cr}\,,
\label{bblock}
\ee
where the entries are themselves $\ell\times\ell$ matrices.  On $U_i$
$$
h_b(x)= h_0+h_1(x-\lambda_i) + h_2(x-\lambda_i)^2+\cdots+
h_{k_i-1}(x-\lambda_i)^{k_i-1}
$$
where the coefficients are the sums of the sub-diagonals:
$$
h_0=b^{i}_{k_i1}, \quad h_1=b^{i}_{k_i-1,1}+
b^{i}_{k_i2}, \quad h_2=
b^{i}_{k_i-2,1}+b^{i}_{k_i-1,2}+b^{i}_{k_i3}, \quad\ldots\,.
$$
The following is a consequence of the definition.
\begin{lemma} For $b\in \fg$ and $h\in \fh$, we have $h_{bh}=
h_bh$ and $h_{hb}=hh_b$.
\end{lemma}
\begin{proposition} Suppose that $a=x$.  If (\ref{alphbet}) holds then
  $a'\in \fh$ and $[a',h_b]=0$. Conversely, suppose that
$h_b(\lambda_i)$ has distinct eigenvalues for each $i$.
If $a'\in\fh$ and $[a',h_b]=0$, then there
  exists $b'\in \fg$ such that (\ref{alphbet}) holds.\label{algprop}
\label{existprop}\end{proposition}
\begin{proof}  
With $a=x$, we have $[a,h]=0$ for all $h\in \fh$. Conversely, if
  $[a,a']=0$ then $a'\in \fh$.  So the first statement follows from
  the lemma since $[a,h_{b'}]=0$ and therefore $[h_b,a']=0$.
So the converse statement is immediate from the following lemma.
\begin{lemma} Suppose that $a=x$ is in Jordan canonical form, 
  and that $c\in\fg$. Then there exists $b\in\fg$ such that $[a,b]=c$ if
  and only if $h_c=0$.
\end{lemma}
\begin{proof}
Suppose that $c$ is given with $h_c=0$.
The off diagonal blocks in $b$ are uniquely determined by those in
$c$, provided that the $\lambda_i$s are distinct.  For a diagonal
block, we must solve
$$
\left[\pmatrix{0&1&0&0 &\ldots&0\cr
0&0&1&0& \ldots&0\cr
0&0&0&1&\ldots&0\cr
\vdots \cr
0&0&0&0& \ldots&0},
\pmatrix{b_{11}&b_{12}&b_{13}&\ldots\cr
b_{21}&b_{22}&b_{23}& \ldots\cr
b_{31}&b_{32}&b_{33}&\ldots\cr
\vdots \cr
b_{k1}&b_{k2}&b_{k3}& \ldots}
\right]=\pmatrix{c_{11}&c_{12}&c_{13} &\ldots\cr
c_{21}&c_{22}&c_{23}& \ldots\cr
c_{31}&c_{32}&c_{33}&\ldots\cr
\vdots \cr
c_{k1}&c_{k2}&c_{k3}& \ldots}
$$
where the $b_{ij}$s and $c_{ij}$s are $\ell\times\ell$ matrices and 
$1$ denotes the identity matrix.  Without loss of generality, we can
set $b_{1j}=0$ for $j=1,\ldots, k$.  By working successively across the
columns of the equation, and by comparing the first $k-1$ entries on
each column on each side, we determine the other entries in $b$.
The equality of the last row on each side then follows from $h_c=0$.
The converse is a consequence of the previous lemma.
\end{proof}
\end{proof}

We want to show that our original deformation problem can be recast in
terms of deformations of the linear system of ODEs (\ref{translinsys}),
where $a\in \fg$ is conjugate to $v \mapsto xv$. To take account of
various `gauge transformations' it is helpful to think of $v$ as lying
not in the fixed vector space $V$, but in a holomorphic bundle $V\to
\cD$ over the deformation manifold.  It is
defined by choosing disjoint open neighbourhoods $U_i$ of the
poles $\lambda_i$ of $A$ and taking the fibre of $V$ at $t\in\cD$ to be the quotient
$$
V_t=\Gamma(\pi^{*}E_t,U)/q\Gamma(\pi^{*}E_t,U)\,,
$$
where $U=\cup U_i$ and $q$ acts by multiplication.  In other words,
the fibre is the set of holomorphic maps $v: U\to E_t$, the modulo maps with
zeros of order $k_i$ at each $\lambda_i$.  A holomorphic section of
$V$ is a section $u$ of $\pi^*E$ over $U$, modulo sections that vanish
to the same order at the $\lambda_i$s.  Our original bundle $E$ is
embedded as sub-bundle of $V$ by mapping sections of $E$ to their
pull-backs by $\pi$.

The 1-form $\alpha$ acts on sections of $V$ by
multiplication:
$$
\alpha:\Gamma(V)\to\Gamma(V\otimes T^*\cD)
:u\mapsto \alpha u\,,
$$
This gives a representation of the algebraic structure on the
tangent space to $\cD$ at each point by endomorphisms of the fibre of
$V$.  The definitions (\ref{abdefn}) of $a,b$ also make sense in this
setting and encode the original connection $\nabla$ in a bundle
endomorphism $b:V\to V$.  Note that $a=i_E\alpha$ and that for any
section $u$ of $V$, the image $bu$ is a section of the subbundle
$E\subset V$. It is unchanged by adding a holomorphic multiple of $q$
to $u$.

If we pick a trivialization for $E$ and introduce the coordinate
$x_i=x-\lambda_i$ on $U_i$, then we can identify the fibres of $V$
with the vector space introduced above.  So a
local trivialization of $E$ gives us a local trivialization of $V$. In
this we have
$$
h_b = Q+O(x_i^{k_i})
$$
at each pole. So we can recover $A$ from $b$.

\subsection*{The connection on $V$}
By assumption, the eigenvalues of $Q$ are distinct at the roots of
$q$, and so by making an appropriate choice for the $U_i$s, we can
assume that this is true throughout $U$. We can then pick holomorphic
sections of $\pi^{*}E\vert_{U}$ which form a basis of eigenvectors of
$Q$ at each point, and so construct another local trivialization of
$V$.  A section $v$ of $\pi^{*}E\vert_U$ is represented by an $\ell
m$-tuple of holomorphic $\C^{\ell}$-valued functions $(v_{i}(x_i))$.
But now $h_b$ is diagonal, with the eigenvalues of $Q$ as diagonal
entries.

In this frame, $a$ is still in Jordan canonical form, while
$\alpha$ takes values $\fa$.
In fact by dropping holomorphic multiples of $q$, we can write
$$
\alpha v\vert_{U_i}= \alpha_iv_i
$$
where $\alpha_{i}$ is a diagonal matrix 1-form on $\cD$, with components
polynomial of degree $k_i-1$ in $x_i$.

The meromorphic connection on $E$ determines a  flat
connection $D$ on $V$, which extends $\nabla_{\infty}$ from the
sub-bundle $E\subset V$ to the whole of $V$.  
It is defined by using the following proposition, in which we use the
new trivialization of $V$.
\begin{proposition} Let $\nabla$ be a meromorphic $q$-connection on
  $\pi^*E$ and define $b$ by (\ref{abdefn}).
Then there is a unique holomorphic connection on $V$ that coincides with
$\nabla_{\infty}$ on $E\subset V$, with the property that
$$
Da=[b,\alpha]+\alpha\,.
$$
\end{proposition}
\begin{proof}
Write $D=\rd + \beta$.  Then the displayed equation gives
\be
[\beta, a] = [b,\alpha]+\alpha-\rd a\,.
\label{betaa}\ee
By Proposition 
(\ref{existprop}), we can solve this for the components of
$\beta$. The solution is not unique since we can add to $\beta$ any
1-form with values in $\fh$.
The freedom is fixed uniquely by imposing the constraint
$(\rd+\beta) v=\nabla_{\infty}v$ for any $v$ independent of $x$.  This
ensures that $D$ coincides with $\nabla_{\infty}$ on sections of $E$.
Note that $b$, and therefore also $\beta$ are {\em not} block diagonal
in this trivialization of $V$.
\end{proof}

\begin{proposition}  $\nabla$ is a solution of the deformation
  problem if and only if $D^2=0$, $Db=0$, and $D\alpha=0$.
\label{Dflat}\end{proposition}
\begin{proof}
  Let $w\in C$ be a point outside the sets $U_i$ and let
  $u$ be a section of $E\subset V$.
  We use the fact that $a$, $\rd a$, and
  $\alpha$ all commute, that $\beta u = \Omega_{\infty}u$, that
  $\beta b =\Omega_{\infty}$ and that $[a,\Omega_{\infty}]=0$.  
With the left-hand side evaluated at $w$,
\begin{eqnarray}
\lefteqn{\bigl(\rd A +\p_w\Omega-[A,\Omega]\bigr)u}\nonumber\\
&=&\Bigl(\rd b -[b,\Omega_{\infty}]-b(a-w)^{-1} (\rd a - \alpha+[b,\alpha])\Bigr)(a-w)^{-1}u\nonumber\\
&=&(\rd b+ [\beta, b])(a-w)^{-1}u 
\label{first}\end{eqnarray}
Similarly, by using $\alpha\vp\alpha=0$,
\begin{eqnarray}
\rd\Omega+\Omega\vp\Omega 
&=&\rd \Omega_{\infty} +\Omega_{\infty}\vp\Omega_{\infty}
+b(a-w)^{-1}(\rd\alpha+\alpha\vp\beta +\beta\vp\alpha)u\nonumber\\
&&\quad
{}+(\rd b +[\beta,b])\vp (a-w)^{-1}\alpha u\,.\label{second}
\end{eqnarray}
Also, from (\ref{betaa}), and by again using $\alpha\vp\alpha=0=\rd
a\vp\alpha=0$,
$$[a,\beta\vp\alpha]=[a,\beta]\vp\alpha =\alpha\vp b\alpha =
-[a,\alpha\vp \beta]\,.$$
Hence $[a,\rd\alpha +\alpha\vp\beta+\beta\vp\alpha]=0$.

Now suppose that $\nabla$ is a solution of the deformation
problem.  Then the left-hand sides of (\ref{first}) and (\ref{second})
vanish, for all $w$.  It follows from (\ref{first}) that $Db=0$; and
from 
(\ref{second}), together with Lemma \ref{vanishing},
that 
$$D\alpha=\rd\alpha+\alpha\vp\beta+\beta\vp\alpha=0\,.
$$
From $Db=\Omega_{\infty}b-b\beta=0$ and we deduce that 
$b(\rd\beta +\beta\vp\beta)=0$.  We also obtain from (\ref{betaa})
that
$$
[\rd\beta+\beta\vp\beta,a]=D\alpha +[b,D\alpha]=0
$$
and hence that $\rd \beta+\beta\vp\beta=0$.

Conversely, suppose that $Db=0$, $D\alpha=0$, and that $D^2=0$.  Then
the curvature of
$\Omega_{\infty}$ vanishes, and so
$$
\rd A +\p_w\Omega-[A,\Omega]=0, \qquad
\rd\Omega+\Omega\vp\Omega 
=0\,,$$
which are the conditions for $\nabla$ to be a solution of the
deformation problem.
\end{proof}

\subsection*{A second trivialization of $V$}

Suppose that $\lambda_i$ is an irregular singularity.  
We can define a diagonal $\ell\times\ell$ 
matrix $w_{i}$, with polynomials in $x_i$ of
degree $k_i-1$ as diagonal entries,  by truncating the Taylor expansion of
$x_it_{i}^{-1/(k_i-1)}$ about $x_i=0$.
That is, \be
w_{i}(x_i)=x_it_{i}^{-1/(k_i-1)} + O(x_i^{k_i})
\label{wdefn}\ee
as $x_i\to 0$.  At a regular singularity, we put $w_i=0$, and note that in any case $w_i=O(x_i)$ as $x_i\to 0$. As
$x\to\lambda_i$
\begin{eqnarray}
g_i^{-1}Ag_i^{\str}\!\!&=&\!\!-\frac{(k_i-1)w_i'}{w_i^{k_i}}+\frac{m_i}{x_i} +O(1)\,,
\label{Aw}\\
g_i^{-1}(\Omega - A\,\rd\lambda_i)g_i^{\str}\!\!&=&\!\!
\frac{(k_i-1)\,\rd w_i}{w^{k_i}}+O(1)
\end{eqnarray}
where the prime is differentiation with respect to $x_i$.

In terms of the matrices $w_i$,
$$
\alpha_i=1_{\ell}\,\rd \lambda_i -(w_i')^{-1}\rd w_i
$$
where the prime again denotes differentiation with respect to $x_i$ and
$\rd$ is the exterior derivative on the deformation manifold, with
$x_i$ held fixed. If we think of $\alpha_i$ as a 1-form on $\cD\times
U_i$, then \be \alpha_i=1_{\ell}(\rd \lambda_i+\rd x_i)-(w_i')^{-1}\rd
w_i\label{alphform} \ee where $\rd$ is now the exterior derivative in
$M\times U_i$.

On $U_i$, the $j$th component of $v$ is holomorphic function $v_{ij}$
of $x_i$.  We get a new trivialization by instead expressing $v_j$ on
$U_i$ as a function of the $j$th diagonal entry $w_{ij}$ in $w_i$. We
note that $w_{ij}$ is a polynoimal in $x_i$, and that it vanishes at
$x_i=0$. So we can represent $v$ by the first $k_i$ coefficients in
the expansion of the $v_j$s in powers of the $w_{ij}$s the new variables on
each $U_i$.

With this choice, $\alpha$ is again a matrix-valued 1-form on $\cD$
with values in $\fa$. As a consequence of (\ref{alphform}) we have
$\rd \alpha=0$ since in the new trivialization of $V$, the exterior
derivative of the $j$the diagonal entry in $\alpha_i$ is taken with
$w_{ij}$ held fixed.
However, 
$a$ is no longer in Jordan form: its diagonal blocks are now
$$
\pmatrix{ \lambda_i & a_{i1} &a_{i2} & a_{13} &\ldots &
  a_{i,k_i-1}\cr
0&  \lambda_i & a_{i1} &a_{i2} & \ldots &
  a_{i,k_i-2}\cr
&& \ddots &\ddots &&&\cr
0&0&0&0& \ldots & \lambda_i\cr           }\,,
$$
where the entries $a_{ij}$ are as in Appendix A, eqn (\ref{aisj});
that is, they are the alternative deformation parameters.  In fact, we
can also read off from (\ref{alphform}) that
$$
\alpha= \rd a
$$
in this trivialization.

The only freedom is in the choice of the
eigenvectors of $Q$ in a neighbourhood of the poles.  If we fix the
order of the eigenvalues, then the trivialization of $V$ is fixed up to
rescaling the eigenvectors of $Q$ by holomorphic functions on $U$.
These leave the
matrix representations of $a$ and $\alpha$ unchanged.
Our matrices $w_{i}$ are determined up to holomorphic multiples of $q$ 
on the open sets $U_i\subset\CP_1$ by the 1-form $A\,\rd
x$ and do not depend on the choice of the coordinate $x$ on the
Riemann sphere.  If we represent $\alpha$ as a 1-form
on $M$ with values in the functions on $U$, and replace $z$ by $tz$,
for some nonzero $t\in \C$, then $\alpha$ is replaced by $t\alpha$.
In the second local trivialization, the diagonal blocks in $\alpha$ are 
all scaled by $t$ (although in the first they transform in a more
complicated way) and so
$$
\Lie_E\alpha=\alpha\,,
$$
where $E$ is the Euler vector field.  Since we also have $\rd
\alpha=0$, we conclude that $\alpha=\rd a$; and also that the flow of
$E$ on $\cD$ is given by the multiplication action of $\C^{*}$ on $a$.
To summarize, we have the following.
\begin{proposition} Locally, in a neighbourhood of a point of the
  deformation manifold at which the eigenvalues of $Q$ are distinct
  and nonzero, there exists a frame for $V$ in which $\rd\alpha=0$,
  $\alpha=\rd a$, and $E(a)=a$.  The frame is determined up to
  transformations which leave the matrix representations of $a$ and
  $\alpha$ unchanged.
\label{closedform}
\end{proposition}

\subsection*{Other forms of the deformation equations}

If we choose a local trivialization of $V$ as in Proposition
\ref{closedform}, then $a$
and $\alpha$ are represented by block diagonal matrices and 
$\alpha=\rd a$.  
We can use the entries in $a$ as local coordinates on
$M$, with the transformation to the deformation parameters used by Jimbo
{\em et al} as in (\ref{coordtrans}).  The matrix representation of $a$ and
$\alpha$ is the same for all local trivializations in this
class; so the coordinates on $M$ are canonical, up to the ambiguity in
taking fractional powers to determine the matrices $w_{i}$ from the
eigenvalues of $A$. 

With this choice of `gauge', the deformation equations become
\be
\rd b +[\beta, b]=0, \qquad [\beta,a]=[b,\rd a]
\label{secform3}\ee
for $b$ as a function of $a\in\fa$, subject to 
the two constraints.  The first is that $b^2=0$.  The second is the
following necessary and sufficient condition that the second equation should be
soluble for $\beta$, given $b$, $a$, and $\rd a$.  
\begin{quote}
(C) Let $g\in\GL$ be such that $gag^{-1}$ is in Jordan
canonical form.  Then 
$$\bigl[h^{\str}_{gbg^{-1}},c\bigr]=0$$ for every $c\in\fa$.
\end{quote}
See Proposition \ref{algprop}. Note that when $a$ is in Jordan
canonical form, it coincides with $v\mapsto xv$.

An alternative route to this form of the deformation equation is to
note that, from  the proof of Proposition \ref{Dflat},
$$
\rd \alpha +\alpha\vp\beta +\beta\vp\alpha=0, 
\qquad \rd \beta+\beta\vp\beta=0\,.
$$
The gauge transformation in Proposition \ref{appprop} in the Appendix 
B then reduces the equations to the form (\ref{secform3}).

\medbreak
\noindent {\bf Example}
The
  Schlesinger equations give the isomonodromic deformations of the
  system
$$
\frac{\rd y}{\rd x}=\sum_1^p\frac{A_iy}{x-\lambda_i}
$$
where the $A_i$s are $\ell \times\ell $ matrices, independent of
$x$.  The deformation parameters are the $\lambda_i$s.  We take $\sum
A_i=0$, so there is no singularity at infinity.
In the alternative formulation, $N=\ell p$ and 
$\fa$ is the set of $N\times N$ diagonal matrices
with diagonal blocks $\lambda_i1_{\ell}$.  We obtain the deformation
equations from (\ref{secform3}) by taking
$$
b=\pmatrix{
A_1&A_2&A_3&\ldots &A_p\cr
A_1&A_2&A_3&\ldots &A_p\cr
\vdots\cr
A_1&A_2&A_3&\ldots &A_p\cr}
$$
The constraint (C) is vacuous in this case, and, 
for example, the $\rd \lambda_1$
component of $\beta_1$
$$
\beta_1=\pmatrix{
\sum_2^p C_{ii}& -C_{22}&-C_{33}& \ldots &C_{pp}\cr
-C_{12}&C_{12}&0&\ldots &0\cr
-C_{13}&0&C_{13}&\ldots &0\cr
\vdots\cr
-C_{1p}&0&0&\ldots &C_{1p}\cr}\,,
$$
where $C_{ij}=A_i/(\lambda_1-\lambda_j)$.
Eqn (\ref{secform3}) then gives
$$
\p_1A_1=\left[A_1,\sum_2^p\frac{A_i}{\lambda_i-\lambda_1}\right],
\quad \p_1A_j=\frac{[A_1,A_j]}{\lambda_i-\lambda_1} \quad (j\neq 1)\,.
$$
The other Schlesinger equations follow similarly. $\Box$
\bigbreak

We shall now look at the deformation equations from three other points
of view.
\begin{itemize}
\item As a Hamiltonian system; 
\item As a symmetry reduction of a generalized form of the
  self-dual Yang-Mills equations; and

\item As equivariance condition on a vector bundle over a `twistor space';
\end{itemize}

\subsubsection*{Hamiltonian system}

The linear system
\be
\frac{\rd y}{\rd z}+\left(a+\frac{b}{z}\right)y=0\,,
\label{linsyst}\ee
where $y$ takes values in $\C^N$, has two singularities---a
regular one at the origin, and a double pole at infinity.  Since the
singularities can be returned to these positions by a M\"obius
transformation, the only non-trivial isomonodromic deformations are
those that change $a$ and $b$.

To understand the Hamiltonian nature of the deformations, we work in
the more general setting of `affine' coadjoint orbits of the loop
group $\LGL$, taking some ideas from Pressley and Segal (1986), pp
49--50.\footnote{The coadjoint orbits of a finite-dimensional Lie
  group are symplectic manifolds; the affine coadjoint orbits appear
  in an extension of this theory due to Souriau in which a translation
  term determined by a Lie algebra cocycle is added to the coadjoint
  action on the dual Lie algebra (see Woodhouse 1990, pp.\ 55-6). In
  this context, the cocycle is given by integrating $\tr (h\rd h')$
  around $\Gamma$ and the affine action is the action of $\LGL$ by
  gauge transformations.}  Let $\Gamma\subset\C$ denote unit circle in
the complex plane, and let $U_+$ and $U_-$ denote the discs on the
Riemann sphere bounded by $\Gamma$, and containing, respectively, the
origin and the point at infinity.
Consider the set of smooth 1-forms on $\Gamma$ with
values in $\gl$.  Each 1-form $$\mu=B\, \rd z$$
determines a monodromy
matrix, up to conjugacy: we pick a fundamental solution
$y:\Gamma\to\GL$ of the differential equation \be \rd y+ \mu y=0
\label{diffeqn2}\ee
and define the monodromy to be the constant matrix $m=y^{-1}\tilde y$,
where $\tilde y$ is the continuation of $y$ once around $\Gamma$ in a
positive sense (see Pressley and Segal 1986, p.\ 124). Since $y$ is
unique up to multiplication on the right by a constant matrix, the
monodromy is determined by $\mu$ up to conjugacy.  If $m=\exp(-2\pi \ri b_0)$ 
for $b_0\in \gl$, then we can construct an element $f$ of the loop group 
$\LGL$ from $y$ amd $m$  by putting $f=y\exp (b_0\log z)$.

We denote by $\cM$ the set of 1-forms for which the monodromy is
conjugate to a fixed matrix $m=\exp(-2\pi\ri b_0)$ and  $f$ lies in the
identity component $\LGL_0$ of the loop group $\LGL$.  
Any two elements $\mu, \hat \mu$ of
$\cM$ are related by a gauge transformation
$$
\mu\mapsto\hat \mu= g^{-1}\mu g+g^{-1}\rd g \,,
$$
where $g$ is found by choosing fundamental solutions $y$ and $\hat y$ of
the corresponding differential equations with monodromy $m$ and by putting
$g=y\hat y^{-1}$. We can
recover $B$ from $f$ by
\be
B\,\rd z=-\rd f f^{-1}+fb_0f^{-1}\,.
\label{Af}\ee
Since $B$ is unchanged when $f$ is multiplied on the right by a
constant matrix that commutes with the monodromy generator, 
we have the following.
\begin{proposition} The space of smooth 1-forms  $\mu$ on $\Gamma$
  with values in $\gl$ and with monodromy conjugate to $m$
  is $\LGL_0/G_0$, where $G_0$ is isomorphic to the
  subgroup of $\GL$ that stabilizes $b_0$ under the
  adjoint representation.  The tangent space to $\cM$ at $\mu$ is
  given by the infinitesimal gauge transformations $\delta \mu=\rD h$
  \be \rD h= \rd h+[\mu,h]\,,
\label{infingauge}\ee
where $h:\Gamma\to \gl$.  Two infinitesimal gauge transformations
determine the same tangent whenever their difference satisfies $\rD
h=0$.
\end{proposition}  
The symplectic form on $\cM$ is defined by
\be
\omega(h,h')=\frac{1}{2\pi\ri}\oint\tr (h\rD h')\, .
\label{symform}\ee
It is skew symmetric, by integration by parts, and non-degenerate
since $\omega (h, \,.\,)=0$ only if $\rd h -\mu h= 0$.  It is closed
since its pull-back to $\LGL$ is $\rd\theta$, where
$$
\theta(h)=-\frac{1}{2\pi \ri}\oint \tr (\mu h)\,.
$$
Note that $\theta$ itself does not descend to $\cM$.

On a dense open subset of $\cM$, we have the Birkhoff factorization
$$
f=f^{-1}_-f_+^{\str},
$$
where $f_+,f_-$ lie respectively in $\LGLp$ and $\LGLm$; that is
the subgroups of loops that extend respectively to holomorphic maps
$U_{\pm}\to \GL$. The factorization is fixed uniquely by
imposing the condition $f_-(\infty)=1$.  From
(\ref{Af}),
$$
f_-^{\str}\mu f^{-1}_--\rd f^{\str}_-f_-^{-1}=
z^{-1}f_+^{\str}b_0f^{-1}_+\,\rd z-\rd
f^{\str}_+f_+^{-1}
$$
The right-hand side is a 1-form on $U_+$, which is holomorphic
apart from a simple pole at 
$z=0$.  We denote it by $\mu_+$ and deduce from the equality that
$$
\mu= f_-^{-1}\mu_+f_-^{\str} +f_-^{-1}\,\rd f^{\str}_-\, .
$$
That is, $\mu$ is gauge equivalent to $\mu_+$.

The only choice is of $f$, which we are free to multiply on the right
by a constant matrix that commutes with $m$.  The effect is
to multiply $f_+$ on the right by the same constant matrix.  Modulo this
freedom, we can parameterize (most of) $\cM$ by $f_-\in \LGLm$, with
$f_-(\infty)=1$, and $\mu_+$. The points that are not
covered are those at which the Birkhoff factorization fails.

If we vary $\mu$, then
\begin{eqnarray*}
\delta \mu \!\!&=&\!\!f_-^{-1}(\delta \mu_+
+[\mu_+,h_-]
+\rd h_-)f_-^{\str},\\
h\!\!&=&\!\!-\delta f\,f^{-1}=f_-^{-1}(h_--h_+)f_-^{\str}\,,
\end{eqnarray*}
where $h_{\pm}=\delta f_{\pm}\,f_{\pm}^{-1}$. We deduce that the
symplectic form can also be written
\be
\omega(h,h')=\frac{1}{2\pi \ri}\oint
\tr(h_-\delta'\mu_+ -h'_-\delta\mu_+-\mu_+[h_-,h'_-] -h_+\delta' \mu_+)\,.
\label{symplus}\ee
Note that $h_-(\infty)=0$.

The isomonodromic deformation equations are obtained by considering
the relationship between two actions on $\cM$.

\subsection*{The loop group and the Marsden-Weinstein reduction}
The identity component of the loop group $\LGL_0$ acts on $\cM$ by
gauge transformations.  If $h$ is a fixed element of the Lie algebra
$\lgl$, which is the set of smooth maps $\Gamma \to \gl$, then its
flow is given by (\ref{infingauge}). It is generated by the function
\be f_h=-\frac{1}{2\pi\ri}\oint \tr(\mu h)\,,
\label{moment}\ee
on $\cM$.  However the action is not Hamiltonian since the Poisson
bracket $\{f_h,f_{h'}\}$ of two generators differs from $f_{[h,h']}$
by a nontrivial Lie
algebra cocycle
$$
\{f_h,f_{h'}\}=f_{[h,h']}+\frac{1}{2\pi\ri}\oint \tr(h'\rd h)\,.
$$
But (\ref{moment}) {\em is} a moment for any subgroup for
which the cocycle vanishes.  One such is
the subgroup $\LGLmtwo$ of loops that extend
holomorphically to $U_-$ and that 
are of the form $1+O(z^{-2})$ as $z\to\infty$.
In this case we get a moment map $\rho$ from $\cM$ into the dual Lie
algebra by putting
$$\langle\rho(\mu),h\rangle=f_h(\mu)\,.$$ 
We note that $\rho(\mu)=0$ if
and only if $B$ extends holomorphically to the outside of $\Gamma$ on
the Riemann sphere, that is, if $\mu$ extends holomorphically except for
a pole of order $2$ at $z=\infty$. The corresponding Marsden-Weinstein
reduction
$$M=\rho^{-1}(0)/\LGLmtwo$$ is a finite-dimensional symplectic
manifold.  Its points are equivalence
classes of holomorphic
maps $B:U_-\to \gl$, which extend smoothly to the boundary $\Gamma$,
with  $B,B'$ equivalent whenever
$$
B\,\rd z=g^{-1}B'g\,\rd z+g^{-1}\rd g,
$$
for some $g:U_-\to \GL$ such that
$g=1+O(z^{-2})$ as $z\to
\infty$.
The action of $\LGLm$ descends to the reduction, as does the
action of the subgroup $\LGLmone$ of loops for which $g(\infty)=1$. 

For any $\mu\in \rho^{-1}(0)$, we can replace the integral around
$\Gamma$ in (\ref{symform}) by any contour in $U_-$ winding once
around the point at infinity. In this case, if $f$ can be factorized, 
then $\mu_+$ holomorphic in $U_-$
apart from a pole of order $2$ at infinity.  Since $\mu_+$ is also
holomorphic on $U_+$ we must have 
$$
\mu_+=(a+z^{-1}b)\, \rd z\, ,
$$
where $a$ is independent of $z$ and $b=kb_0k^{-1}$, with $k=f_+(0)$. 
So apart from the singular points at which the factorization fails, 
points of $M$ are parametrized by $a,b$,
together with $f_-\in\LGL_-$ of the form
$$
f_-=1+z^{-1}p
$$
Evaluation of the residues in (\ref{symplus})
then gives the reduced symplectic form as
$$
\omega=\tr\bigl(\rd p\vp \rd a-b_0k^{-1}\,\rd k\vp k^{-1}\,\rd k\bigr)\,,
$$
where $k=f_+(0)$. 
Thus $a$ and $p$ are conjugate variables, while the last term is
the natural symplectic form on the adjoint orbit of $b_0$---the orbit
determined by the monodromy.  The symplectic form coincides with that on
$T^{*}\fg\times \cO$, where $\cO$ is the adjoint orbit
of the monodromy generator $b_0$, and $\fg=\gl$.

\subsection*{Flows of spectral invariants}
These flows are generated by Hamiltonians constructed from the
eigenvalues of $B$.  They are defined in a neighbourhood of
$\Sigma\subset M$, where $\Sigma$ is the subset of $M$ 
defined by the
condition that $a$ should be conjugate to an element of $\fa$.

Fix, for the moment, a base point $\mu=B\,\rd z\in \Sigma$. For generic $b$,
the set of eigenvalues of $B$ near $z=\infty$ is partitioned into subsets labelled by the Jordan blocks in the canonical form of $a$.   
As
$z\to \infty$, the eigenvalues corresponding to block $J$ are of the form
\be
\nu =\lambda +O(z^{-1/k})
\label{asym}\ee
where $\lambda$ is an eigenvalue of $a$ of $J$ and $k$ is the size of $J$.  
A circuit of $\infty$ permutes the eigenvalues within each partition 
but does not mix the eigenvalues in different partitions. Near $z=\infty$, 
we can write
\be
B=g(z)\Delta g(z)^{-1}\,,
\label{diag}\ee
where $\Delta$ is diagonal, with the $\nu$s on the diagonal.  The
matrix $g(z)$ is holomorphic near $\infty$, except that it is not
single-valued and has a branching singularity at $\infty$.  However
$g(z)=o(z)$ as $z\to \infty$.  If $\mu'=B'\,\rd z$ is close to $\mu$,
then its eigenvalues near $z=\infty$ can be partitioned in the same way,
according their limiting behaviour as $B'\to B$ for fixed $z$.

The Hamiltonians in which we are of two types. The first type are
\be H_{i}=\frac{1}{4\pi\ri}\oint z\sum \nu^2\,\rd z
\label{Hi}\ee
where $i$ labels the poles of $A$ in the original problem and 
the sum is over all the blocks for which $\lambda=\lambda_i$.
The second is
$$ 
H_{Jj}=\frac{1}{2\pi \ri (j+1)}\oint z \sum_J
(\nu-\nu_0)^{j+1}\,
\rd z, \qquad
j=1,\ldots, \vert J\vert,
$$
where $\vert J\vert$ is the size of block $J$ and $\nu_0=\sum_J\nu$.  

It is claimed that the flows of both type of Hamiltonian are tangent
to $\Sigma$.  To establish this, we find the variations $\delta_iB$
and $\delta_{Jj}B$ generated by the Hamiltonian flows at the base point. We
note first that that these are orthogonal with respect to $\omega$ to
all variations $\delta B$ that leave $g(z)$ unchanged.
We can construct other variations $\delta_{\xi_J} B$ by replacing the
eigenvalues in (\ref{diag}) of a block $J$ 
by the corresponding branches of a
convergent power series
$$
\xi_J= \xi_0 + \xi_1z^{-1/k} + \xi_2 z^{-2/k} +\cdots
$$
in $z^{1/k}$, and setting the other entries in $\Delta$ to zero.
The result is a variation $\delta B$ which is holomorphic in a
neighbourhood of $z=\infty$, since $\delta B$ is single-valued on
$\Gamma'$ and equal to $o(z)$ as $z\to\infty$.  The variations
generated by the Hamiltonians at the base point are determined by
their $\omega$-inner products with these variations.

Let $B_J$ denote the matrix obtained by taking $\xi_J=\nu$.  
$$ 
H_{Jj}=\frac{1}{2\pi \ri (j+1)}\oint\tr (z (B_J-\tr B_J)^{j+1})\rd
z\, .
$$
Under a variation given by $\xi_{J'}$ for $J'\neq J$, we have
$\delta H_{Jj}=0$.  While for a variation given by $\xi_J$, we have
$$
\delta \tr(B_J-\tr B_J)^{j+1}
=(j+1)\tr\bigl((B_J-\tr B_J)^j\,\delta_{\xi_J} B\bigr)\,.
$$
Since we can evaluate $\omega$ at the base point by using any small
circuit of $z=\infty$, we conclude that the Hamiltonian flow generated
by $H_{Jj}$ is given at $\mu$ by
$$
\delta_{Jj} B= \rD (z(B_J-\tr B_J)^j)\,.
$$
This preserves the canonical form of $a$ and is therefore tangent to $\Sigma$
We similarly find the flow of $H_i$ to be
$$
\delta_j B= \sum \rD (z B_J)
$$
with the same sum as in (\ref{Hi}) and
The $H_i$s are in involution, while
$$
\{H_{Jj},H_{J'j'}\}=\left\{\begin{array}{ll}0 \quad& \mbox{if $c\neq
      c'$}\\
(j-j')H_{J,j+j'-1}&\mbox{if $J=J'$.}\end{array}\right.
$$
For $j>\vert c\vert$, $H_{Jj}$ is defined by the same formula, but
the Hamiltonian vector field is linearly dependent on those for lower
values of $j$.

\begin{proposition} The flows
of the Hamiltonians $H_{Jj}$ map orbits of $\LGLmone$ in $\Sigma$ 
to orbits of $\LGLmone$.
\end{proposition}

\begin{proof} Let
  $\mu \in\rho^{-1}(0)$ be the representative of a point in $\Sigma$, 
so that $B$ and the matrices $B_J$ extend
  holomorphically to $D_-$.  The infinitesimal flow of $H_{Jj}$ is
  $$
  \delta_{Jj}B=[h_{Jj},B]-\p_zh_{Jj}=-\p_zh_{Jj}\,,$$
  where $h_{Jj}=z(B_J-\tr)^j$. The right-hand side is holomorphic at
  infinity, so the flow at $\mu$ is tangent to $\rho^{-1}(0)$.

Let $h\in\lglm$, the Lie algebra of $\LGLm$.  This
  generates the flow $\delta_hB=\rD h$.  Therefore
$$
\delta_hB_J=[h,B_J]+O(z^{-2})
$$
as $z\to \infty$. Since $\delta_{Jj}k=0$, we have
$$
\delta_k(h_{Jj})-\delta_{Jj}k-[h_{Jj},k]=O(z^{-1})$$ as
$z\to\infty$. Therefore the commutator of the two flows coincides at
$B$ with the flow of a generator of $\LGLmone$.
\end{proof}

How do the flows of the Hamiltonians give solutions of the isomonodromy
problem?  Consider the submanifold $\Sigma_+ \subset \Sigma$  given by
$\mu=\mu_+$.  This is $p=0$ in our parameterization.  For nearby
points $B$ at which the  Birkhoff factorization exists, we have a
projection the $B\mapsto B_+$ onto $\Sigma_+$, given by setting $p=0$.
Since the flow of $H_{Jj}$ maps orbits of $\LGLmone$ to orbits of
$\LGLmone$, the flows project onto $\Sigma$. If in the notation of the
proof, we write
$$
h_{Jj}=z\alpha_{Jj}+\beta_{Jj}+O(z^{-1})\,,
$$
then, since $[B,h_{Jj}]=0$, we have
$$
[a,\alpha_{Jj}]=0, \qquad [a,\beta]=[\alpha_{Jj},b]
$$
and the projected flow on $\Sigma$ is
$$
\delta_{Jj} \,a =\alpha_{j,J}, \qquad  \delta_{Jj}\,b+[\beta_{Jj},b]=0\,.
$$
In particular, when $a\in \fa$, then the flows give the solutions
of (\ref{secform}).  The picture is similar for the $H_i$s.  The
deformations generated by the $H_i$s move the positions of the poles
in the original deformation problem without changing the behaviour at
$A$ at is poles beyond a local coordinate transformation.  Those
generated by the $H_{Jj}$s fix the poles of $A$, but change the other
deformation parameters.

\subsection*{`Self-dual' connections} 

We now turn to the interpretation of (\ref{sdconn}) as a `self-dual'
connection.  In the standard theory, the Penrose-Ward transform
identifies solutions of the self-dual Yang Mills equations in regions
of complex Minkowski space (or in Euclidean space) with holomorphic
vector bundles over corresponding open sets in the complex projective
space $\CP_3$.  The statement and proof of the theorem (Ward and Wells
1990, Theorem 8.1.2) carry over almost unaltered to the present
context.  The only difference is that the twistor space is now
$\P(\fa\oplus\C^2)$ instead of $\CP_3$. The variables $\omega\in\fa$
and $\pi_A\in\C^2$ are homogeneous coordinates on $\cZ$, and each null 
$N$-plane determines a point in $\cZ$, through (\ref{nullplane}).
Conversely, every point of $\cZ$ determines a unique null plane,
except the points on the line $I=\{\pi_A=0\}$.  As in the standard
theory, we can compactify $\M$ by adding points at
infinity to include these exceptional null $N$-planes, but that is not
explored here.

Going in the other direction, each point in $\M$ gives a 
line in $\cZ$, by reading (\ref{nullplane}) the other way around, with
$s,t$ fixed and $\omega$ and $\pi_A$ varying. Conversely, every line in
$\cZ$, other than those that intersect $I$, determines a point in
$\M$. The lines that intersect $I$ correspond to the points at infinity
in the compactification of $\M$.

The correspondence maps self-dual connections on the trivial vector
bundle over a convex region $W\subset \M$ to holomorphic bundles over
over $\cZ$, where $\cZ$ is the set of null $N$-planes that
intersect $W$. It is one-to-one provided that the bundle
is trivial on one line in
$\cZ_W$, and therefore on the generic line.

In the this context, the role of the conformal group of space-time
falls to the group transformations of (the compactification of) $\M$
that map null $N$-planes to null $N$-planes.  These include the three flows
$$
(s,t)\mapsto (s,\re^{\tau}t), \qquad
(s,t)\mapsto (s+\tau c, t), \qquad
(s,t)\mapsto (s, t+\tau c)\,,
$$
where $\tau\in\C$ is the parameter along the flows and $c\in\fa$ is constant.
We denote the respective generating vector fields on $\M$ by $E$,
$S_c$, and $T_c$.  Since the flows map null $N$-planes to null $N$-planes,
they induce flows on twistor space, given respectively by
$$
(\omega,\pi)\mapsto (\omega,\re^{-\tau}\pi_0,\pi_1), \quad
(\omega,\pi)\mapsto (\omega+\tau\pi_1c,\pi),\quad
(\omega,\pi)\mapsto (\omega+\tau\pi_0c,\pi)\,.
$$
We shall also denote the corresponding generating vector fields on
twistor space by $E$, $S_c$, and $T_c$. 

The particular `self-dual' connections that
correspond to solutions of the isomonodromy problem are equivariant
along the flows of $E$ and $S_c$ on twistor space for all $c\in\fa$, 
and satisfy special conditions at
the fixed points of these flows.  I shall not go into the details of
this construction because they follow exactly the same pattern as the 
standard examples explored in Ward and Wells (1990) or in Mason and
Woodhouse (1996): the construction is important here only because it
provides the geometric context of the following more explicit
reduction to a Riemann-Hilbert problem. 
 The transition is obtained by
 introducing the inhomogeneous coordinates 
 $$
 Z=\omega/\pi_1, \qquad
 z=\pi_0/\pi_1\,,
 $$
 and by representing the bundle by its patching matrix $P(Z,z)$. In
 the inhomogeneous coordinates, eqn (\ref{nullplane}) becomes
 $Z=s+zt$, and the flows of $S_c$ and $T_c$ become, respectively,
$$
Z\mapsto Z+\tau c \quad (S_c), \qquad Z\mapsto Z +\tau z c\quad  (T_c)\,,
$$
where $\tau\in\C$ is the parameter along the flow.

\subsection*{Reduction to a Riemann-Hilbert problem}

Let $\cZ$ be an open neighbourhood of a line in $\P(\fa\oplus\C^2)$
and let $\cZ_+$ and $\cZ_-$ be the complements in $\cZ$ of,
respectively, the hyperplanes $z=\infty$ and $z=0$.  We assume that
$\cZ=\cZ_+\cup\cZ_-$; that is, that it contains no points at which
$\pi_A=0$. Let
$$
P:\cZ_+\cap\cZ_-\to \GL
$$
be holomorphic.  By restriction, $P$ determines a map
$P_X:\C\setminus \{0\}\to\GL$ for each line $X\subset \cZ$.  If $X$ is
given by (\ref{nullplane}) and $P$ is expressed as a function of the
inhomogeneous coordinates $Z,z$, then
$$
P_X(z)=P(s+zt,z)\,.
$$
We write the Birkhoff factorization as 
\be
P_X=f_-^{-1}(s,t,z)\Delta f^{\str}_+(s,t,z)\, ,
\label{birk}\ee
where $f_-,f_+$ are holomorphic at, respectively, $z=\infty$ and $z=0$,
and $\Delta$ is diagonal with powers of $z$ on the diagonal.
\begin{proposition}
Suppose that $\Delta=I$ for one in line $\cZ$; and
$S_c(P)=\theta_cP$, $E(P)=-Pb_0$,
where $\theta_c:\cZ_-\to\gl$, with
$\theta_c=c$ on $\pi_1=0$, and $b_0$ is constant.
Then $\Delta=I$ for a dense open subset of the lines in $\cZ$.  The
Birkhoff factorization can be fixed uniquely on this
set by the condition $f_-(s,t,\infty)=\exp(s)$. In this case, $k=f_+(s,t,0)$
is independent of $s$ and 
a solution of (\ref{secform}) with monodromy matrix $m=\exp(2\pi\ri b_0)$  
is obtained by putting
$\beta=k^{-1}\rd k$ and $b=i_E\beta$.
\label{patch}\end{proposition}
The proposition  is a translation into a concrete form of the Atiyah-Ward
correspondence for bundles with the appropriate equivariance.  We
shall prove it directly rather than by working through the details of
the correspondence in this case.
\begin{proof}
The first statement is a standard part of the theory of Birkhoff
factorizations (see Pressley and Segal (1986), Ch.\ 8 ). When $\Delta=I$,
we are free
to transfer a $z$-independent matrix from $f_+$ to $f_-$, so we have
the freedom to impose the `gauge condition'.  
The remaining statements follow by
differentiating (\ref{birk}) and by applying Liouville's theorem.
Since $P$ is constant on $N$-planes, we have for each $c\in\fa$
$$
T_c(f^{\str}_+)f_+^{-1}-zS_c(f^{\str}_+)f_+^{-1}=
T_c(f^{\str}_-)f_-^{-1}-zS_c(f^{\str}_-)f_-^{-1}\,.
$$
The left-hand side is holomorphic in $z$ at the origin, and the
right-hand side has a simple pole at $z=\infty$.  We conclude from
Liouville's theorem that both sides must be equal to
$$
-\beta_c-z\alpha_c
$$
for some $\gl$-valued functions 
$\alpha_c$ and $\beta_c$ on $\M$, depending linearly on $c$. By considering the
behaviour at $z=\infty$ and by using the gauge condition, we have
$\alpha_c=c$.

By differentiating (\ref{birk}) along the flows of $S_c$ on $\cZ$ and $\M$,
$$
S_cP=\theta_cP=-f_-^{-1}S_c(f^{\str}_-)f_-^{-1}f^{\str}_+ +f_-^{-1}S_cf^{\str}_+\,,
$$
from which we obtain
$$
 S_c(f^{\str}_-)f_-^{-1}-f^{\str}_-\theta_cf_-^{-1}=
S_c(f^{\str}_+)f_+^{-1}\,.
$$
The left hand-side is holomorphic and vanishes at $z=\infty$ by
the properties of $\theta_c$ and the gauge condition.  The right-hand
side is holomorphic at $z=0$.  By Liouville's theorem, both sides are
identically zero.  It follows that
$f_+$ and $k$ are independent of $s$
and hence that 
$$
T_c(f^{\str}_+)f_+^{-1}=-\beta_c-zc\,.
$$
We also have
$$
-z\p_z P=Pb_0=-f_-^{-1}E(f^{\str}_-)f_-^{-1}f^{\str}_+
+f_-^{-1}Ef^{\str}_+\,,
$$
and hence
$$
E(f^{\str}_+)f_+^{-1}-f^{\str}_+b_0f_+^{-1}=E(f^{\str}_-)f_-^{-1}\,.
$$
Again both sides must be independent of $z$.  Since the right-hand
vanishes at $z=\infty$, where $f_-$ is independent of $t$, we conclude
that both sides vanish identically.

It follows that if we define a 1-form $\beta$ on $\M$ by 
$$
i_{S_c}\beta=0,\qquad i_{T_c}\beta=-\beta_c\,,
$$
then $\beta=(\rd k)k^{-1}$ and $\Lie_E\beta=0$. The proposition follows.
\end{proof}

Finally we prove Proposition 1.  Suppose that $b(a)$ is a solution of
(\ref{secform})
satisfying the constraints.  Then we the linear equations 
$$
\rd f -\beta(a)f - z\, \rd a f=0
$$
can be solved, with $f$ holomorphic in $z$ at $z=0$.  If we also
have $\Lie_E\beta=0$, then $\beta=\rd k\,k^{-1}$, where $k(a)=f(a,0)$,
and $b=kb_0k^{-1}$ for some constant $b_0$.  By exploiting the homogeneity
of the linear system under the flow of $E$, we can ensure that
$$
f(\lambda a, \lambda^{-1} z)=f(a,z), \qquad \lambda\in\C^{*}\,.
$$
It then follows that
$$
\p_zf +(a+z^{-1}b)f = z^{-1}fb_0
$$
and consequently that $y=f\exp (b_0\log z)$ is a solution to the linear
system (\ref{translinsys}).  Since this system has a regular singularity at
$z=0$, and no other singularity at finite values of $z$, it follows
that $f$ is holomorphic for all $z$. Proposition 1 now follows by
taking $P(Z,z)=f(z^{-1}Z,z)$ and by applying Proposition (\ref{patch}).

\subsection*{Appendix A}

Here we find the transformation between the orginal parameters
$\lambda_i, t_i$ on the deformation manifold and the entries in $a$.

At an irregular singularity $\lambda_i$, 
let $w_i$ be the matrix defined in (\ref{wdefn}).
At a regular singularity, we put $w_i=0$. Note that
in any case $w_{i}(0)=0$ and that
the polynomial coefficients of $w_{i}$ depend on those of $t_{i}$
but not on the exponents of formal monodromy nor on the positions of the poles.
At $x=\lambda_i$,
$$
g_i^{-1}Ag_i=\frac{\rd}{\rd x_i}\left(\frac{1}{w_i^{k_i-1}}\right)
+\frac{m_i}{x_i} + O(1)\, \qquad \mbox{as $x_i\to 0$.}
$$
By inverting the relationship between the entries in $w_i$ and
$x_i$, we can write 
\be x_i1_{\ell}=a_{i1}w_{i} + a_{i2}w_{i}^2 +
\cdots +a_{i,k_i-1}w_i^{k_i-1}\,,
\label{aisj}\ee
where the coefficients $a_{ij}$ ($i=1, \ldots, m$, $j=1,
\ldots,k_i-1$)
are diagonal matrices with complex
entries.
Since
$$t_i(a_{i1}w_{i} + a_{i2}w_{i}^2 +
\cdots\quad )= (a_{i1} + a_{i2}w_{i} +a_{i3}w_{i}^2
+\cdots\quad)^{k_i-1} + O(x_{i}^k).$$
we can relate the $a_{ij}$s to the matrices $t_i$ by 
comparing coefficients of powers of $w_{i}$.  The result is that
the coefficients of the diagonal entries in $t_i$ are polynomials
in the diagonal entries in the $a_{ij}$s:
\begin{eqnarray}
t_i&\!\!=\!\!&a_{i1}^{k_i-1}
+(k_i-1)a_{i1}^{k_i-3}a_{i2}x_i
+\Bigl(\half (k_i-1)(k_i-4)a_{i1}^{k_i-5}a_{i2}^2+(k_i-1)a_1^{k_i-4}
a_{i3}\Bigr)x_i^2 +
\nonumber\\
&&\Bigl(\sixth (k_i-1)(k_i-5)(k_i-6)a_{i1}^{k_i-7}a_{i2}^3+
(k_i-1)(k_i-5)a_1^{k_i-6}a_{i2}a_{i3} +
(k_i-1)a_{i1}^{k_i-5}a_{i4}\Bigr)x_i^3\nonumber\\
&&{}\quad +\quad\cdots \label{coordtrans}
\end{eqnarray}
The coefficients of terms with negative powers of $a_{i1}$ all vanish.
It is the diagonal entries in the $a_{ij}$s that we shall use in place
of the parameters in Jimbo {\em et al} as coordinates on $\cD$---or
more precisely on a covering space of $\cD$.

\subsection*{Appendix B}

Suppose that $\alpha$ and $\beta$ are $\GL$-valued 1-forms such that 
$\alpha$ takes values in $\fa$ and
$$
\rd \alpha +\alpha\vp\beta +\beta\vp\alpha=0, \qquad \rd \beta+\beta\vp\beta=0\,\qquad \Lie_E\alpha=\alpha.
$$
We assume that $i_E\alpha$ has the generic Jordan
canonical form of the elements of $\fa$.

We consider gauge transformations the form
$$
\alpha \mapsto h^{-1}\alpha h
\qquad 
\beta\mapsto \rd h \, h^{-1} +h\beta h^{-1}\,.
$$
\begin{proposition} 
  The gauge can be chosen locally so that $\rd \alpha=0$,
  $\Lie_E\alpha=\alpha$, and $\alpha$ takes values $\fa$.
  \label{appprop}
\end{proposition}

\begin{proof}
  We consider first the case in which the matrices in $\fa$ have only
  one Jordan block of size $n$.  We denote by $\fk$ the normalizer of $\fa$ in
  $\gl$. This is spanned by the matrices of the form
$
a+Da'
$
where $D={\rm diag}\, (0,1,2,,3, \ldots\:)$ and 
$a,a'$ are in $\fa$; that is, $a,a'$ are of the form
$$
a=\pmatrix{a_0&a_1& a_2& a_3&\ldots \cr
0&a_0& a_1&a_2 \cr
0&0&a_0& a_1\cr
&&\ddots&\ddots&\ddots\cr}\,.
$$
We note that $[Da,Da']=Da''$, where
$a''=[a',D]a-[a,D]a'$.
\begin{lemma} There exists a 1-form 
$\xi$ with values in $\fk$ such that 
\be
\rd\alpha+\alpha\vp \xi
+\xi\vp \alpha=0, \qquad i_E\xi=0\, .
\label{kapalph}\ee
\end{lemma}
\begin{proof}
Write $\beta=(\beta_{ij})$, and
$$
\alpha=\pmatrix{\alpha_0&\alpha_1& \alpha_2& \cr
0&\alpha_0& \alpha_1&\ddots \cr
&&\ddots&\ddots}\,.
$$
By the assumption that $a$ has the generic Jordan canonical form,
$a_1=i_E\alpha_1\neq 0$.  We shall look for $\xi$ of the form
$$
\xi=D
\pmatrix{\xi_0&\xi_1&\xi_2& \ldots \xi_{n-2}&0\cr
             0&\xi_0&\xi_1& \ldots &\xi_{n-2}\cr
             0&0&\xi_0&\ldots &\xi_{n-3}\cr
             \vdots &\vdots &\vdots &\ddots& \vdots \cr 
             0&0&0&\ldots &\xi_0}\,,
$$      
with $i_E\xi_i=0$. 
Put $\alpha_i=0$ for $i<0$.  Then eqn (\ref{kapalph}) reads
\be
\rd\alpha_{j-i}
+\sum_{k=1}^n\bigl(\alpha_{k-i}\vp \beta_{kj}+\beta_{ik}\vp\alpha_{j-k}\bigr)=0\, .
\label{gamma2}\ee
By putting $i=j$ and summing over $i$,
we find $\rd\alpha_0=0$.

Given $\beta$,
therefore, the problem is to find $\xi_0, \ldots, \xi_{n-2}$
such that
\begin{eqnarray}
&&\rd\alpha_1+\alpha_1\vp \xi_0=0\nonumber\\
&&\rd\alpha_2+\alpha_1\vp \xi_1+2\alpha_2\vp \xi_0=0\nonumber\\
&&\rd\alpha_3+\alpha_1\vp \xi_2+2\alpha_2\vp \xi_1+3\alpha_3\vp \xi_0=0\nonumber\\
&&\qquad\vdots\nonumber\\
&&\rd\alpha_{n-1}+\alpha_1\vp \xi_{n-2}+2\alpha_2\vp
\xi_{n-3}+\cdots +(n-1)\alpha_{n-1}\vp \xi_0 =0
\label{kapalph1}\end{eqnarray}
If $\xi_0, \ldots ,\xi_{r-3}$ are known, and the first $r-2$
equations hold, then $\xi_{r-2}$ can be found so that the $(r-1)^{\rm
th}$ equation also holds, provided that
$$
\alpha_1\vp\Bigl(\rd\alpha_{r-1}+2\alpha_2\vp\xi_{r-3} +\cdots +
(r-1)\alpha_{r-1}\vp\xi_0\Bigr)=0\, ;
$$
That is, provided that,
\be
\alpha_1\vp\rd\alpha_{r-1}+2\alpha_2\vp\rd\alpha_{r-2}+3\alpha_3\vp\rd\alpha_{r-3}
+\cdots +(r-1)\alpha_{r-1}\vp\rd\alpha_1=0\, .
\label{algcond}\ee
So we can find $\xi$ provided that this holds for $r=1, \ldots, n-1$.

We can rewrite (\ref{algcond}) as
$$
\sum_{i=p}^n\sum_{j=1}^q\alpha_{i-j+r}\vp\rd\alpha_{j-i}=0
$$
for any $0\leq p,q\leq n$ such that $r=q-p+1$.  By taking the exterior
product of (\ref{gamma2}) with $\alpha_{i-j+r}$ and summing, we obtain
$$
\sum_{i=p}^n\sum_{j=1}^q\alpha_{i-j+r}\vp\rd\alpha_{j-i}
+\sum_{i=p}^n\sum_{j=1}^q\sum_{k=1}^n\alpha_{i-j+r}\vp
\alpha_{k-i}\vp\beta_{kj}+\sum_{k=p}^n\sum_{i=1}^q\sum_{j=1}^n\alpha_{k-i+r}
\vp\beta_{kj}\vp\alpha_{i-j}=0\, .
$$
In the case $m=k-p-q+j>0$, 
the coefficient of $\beta_{kj}$ ($k\geq p$, $j\leq q$) is
$$
\sum_{i=p}^{k-1}\alpha_{i-j+r}\vp\alpha_{k-i}-\sum_{i=j+1}^q\alpha_{k-i+r}\vp
\alpha_{i-j}=\sum_{i=1}^{m}\alpha_{i+q-j}\vp\alpha_{k-p-i+1}=0\, .
$$
Similarly in the cases $m=0$ and $m<0$.  It follows that
(\ref{algcond}) holds.  The 1-form $\xi$ is uniquely determined by
(\ref{kapalph1}) up to the addition of terms in $\alpha_1$.
We use this freedom to impose the additonal condition that
$i_E\xi=0$, which then determines $\xi$ uniquely.
\end{proof}

\begin{lemma} 
$\rd \xi+\xi\vp\xi=0$.
\end{lemma}
\begin{proof}
We find $\xi$, as in the previous lemma.
Put $\Xi=\rd \xi
+\xi\vp\xi$.  Then 
$$  
\Xi=
\pmatrix{0&0&0& \ldots &0\cr
             0&\omega_0&\omega_1& \ldots &\omega_{n-2}\cr
             0&0&2\omega_0&\ldots &2\omega_{n-3}\cr
             \vdots &\vdots &\vdots &\ddots& \vdots \cr 
             0&0&0&\ldots &(n-1)\omega_0}\, ,
$$
where
\begin{eqnarray*}
\omega_0&=&\rd \xi_0\\
\omega_1&=&\rd\xi_1+\xi_1\vp\xi_0\\
\omega_2&=&\rd\xi_2+2\xi_2\vp\xi_0\\
\omega_3&=&\rd\xi_3+3\xi_3\vp\xi_0+\xi_2\vp\xi_1\\
&\vdots&\\
\omega_k&=&\rd\xi_k+\sum_{i>\frac{k}{2}}(2i-k)\xi_i\vp\xi_{k-i}\,
.\\
\end{eqnarray*}
From the definition of $\Xi$ as a curvature form, we obtain
\be
\rd\Xi+\xi\vp\Xi-\Xi\vp\xi=0\, .
\label{dOmega}\ee
By taking the exterior derivative of (\ref{kapalph}), we obtain
$$
\rd\xi\vp\alpha
-\xi\vp\rd\alpha+\rd\alpha\vp\xi-\alpha\vp\rd\xi=0
$$
and hence from(\ref{kapalph}) again
\be
\Xi\vp\alpha+\alpha\vp\Xi=0\, .
\label{Omalph}\ee
By taking the inner product of (\ref{kapalph}) with $E$ and by using
$\Lie_E\alpha=\alpha$, we have
$$
[a,\xi]=\rd a-\alpha\, ,
$$
where $a=i_E\alpha$.
By taking the exterior derivative of (\ref{kapalph}), we have
$$
\rd a\vp \xi+\xi\vp\rd a +[a,\rd\xi]=-\rd \alpha
$$
and hence, by taking the inner product with $E$ again, and by using
$E(a)=a$, we have
$$  
[a,\xi]-\rd a +\alpha +[a,i_E\,\rd \xi]=0\,.
$$
Therefore $[a,i_E\xi]=0$.  The particular form of $\xi$ then gives
$i_E\xi=0$.  It follows that $i_E\Xi=0$ and hence from (\ref{Omalph}) that
$[a,\Xi]=0$.  Finally, again from the special form of $\Xi$, we obtain $\Xi=0$.
\end{proof}
Both lemmas hold for general $\fa$, by applying the same arguments to
each diagonal block. From the second lemma, we have $\xi=-(\rd h)
h^{-1}$
for some $h$ taking values in the normalizer group of $\fa$.  
If we make a gauge transformation by $h$, then
$$
\rd\alpha \mapsto \rd(h^{-1}\rd\alpha h)=
h^{-1}(\xi\vp\alpha +\rd \alpha +\alpha\vp \rd \xi)h=0\, ,
$$
which concludes the proof of the proposition.
\end{proof}

\subsection*{References}

\setlength{\parindent}{0in}
\setlength{\parskip}{0.15in}
Beals, R, and Sattinger, D.H. (1993). Integrable systems and isomonodromy
deformations. {\em Physica} {\bf D65}, 17--47.

Bertola, M., and Mo, M.Y. (2005). Isomonodromic deformation of resonant
rational connections. arXiv:nlin.SI/0510011 v1 5 Oct 2005.

Dubrovin, B. (1996). 
Geometry of 2D topological field theories. In: {\em Integrable
systems and quantum groups}. Ed M. Francaviglia and S. Greco. Lecture
Notes in Mathematics, 1620. Springer, Berlin.

Harnad, J. (1994).  Dual isomonodromic deformations and moment maps to
loop algebras.  {\em Comm.\ Math.\ Phys.} {\bf 166} (1994), 337--365.

Hitchin, N.J. (1997).  Frobenius manifolds.  With notes by David
Calderbank. {\em NATO Adv.\ Sci.\ Inst.\ Ser.\ C Math.\ Phys.\ Sci.},
  {\bf 488}, Gauge theory and symplectic geometry.  Kluwer, Dordrecht.

Jimbo, M., Miwa, T., and Ueno, K. (1981a). Monodromy preserving 
deformations of 
linear ordinary differential equations with rational coefficients. I. General 
theory and $\tau$-functions. {\em Physica} {\bf 2D}, 306--52.

Jimbo, M., and Miwa, T. (1981b). Monodromy preserving deformations of 
linear ordinary differential equations with rational coefficients. II. 
{\em Physica} {\bf 2D}, 407--48.

Jimbo, M., and Miwa, T. (1981c). Monodromy preserving deformations of 
linear ordinary differential equations with rational coefficients. III. 
{\em Physica} {\bf 4D}, 26--46.

Mason, L. J., and Woodhouse, N. M. J. (1996). {\em Integrability, 
self-duality, and twistor theory}. 
Oxford University Press,  Oxford. 

Pressley, A., and Segal, B. (1986). {\em Loop groups}. 
Oxford University Press,  Oxford. 

Ward, R.S., and Wells, R. O. ((1990). {\em Twistor geometry and field theory}. Cambridge University Press, Cambridge.

\end{document}